\documentclass{pramana}

\usepackage{graphicx,amsmath,bm}
\usepackage[super,compress]{cite}
\usepackage{graphics}
\usepackage{graphicx}
\usepackage{float}
\usepackage[colorlinks,citecolor=blue,urlcolor=blue,linkcolor=blue]{hyperref}
\usepackage[lofdepth,lotdepth]{subfig}
\usepackage{epstopdf}

\begin{document}

\title{On Distinguishing Different Models of a Class of \\ Emergent Universe Solutions}


\author{Souvik Ghose\textsuperscript{1}}
\affilOne{\textsuperscript{1} Department of Physics, SIEM, Siliguri\\}


\twocolumn[{

\maketitle

\corres{mrsghose@gmail.com}


\begin{abstract}
A specific class of singularity free cosmological model has recently been considered in light of different observational data like Observed Hubble Data, BAO data from Luminous Red Galaxy survey by Slowan Digital Sky Survey (SDSS) and CMB data from WMAP. However it is observed that only $12-14$ data points are used to study the viability of the model in late time . In this paper we discuss the viability of all the models belonging to the same class of EU in light of Union Compilation data (SnIa) which consists over a hundred data points, thus getting a more robust test for viability. More importantly it is crucial that we can distinguish between the various models proposed in the class of solution obtained. We discuss here why with present observational data it is difficult to distinguish between all of them. We show that the late time behaviour of the model is typical to any asymptotically de-Sitter model.

\end{abstract}


\newcommand{\etal}{{\it et al.}}


}]


\doinum{12.3456/s78910-011-012-3}
\artcitid{\#\#\#\#}
\volnum{123}
\year{2016}
\pgrange{23--25}
\setcounter{page}{23}
\lp{25}

\section{Introduction}
\label{intro}

The present phase of accelerated expansion of the universe \cite{riess, perlm, perlm2, perlm3} seems to be an undeniable fact today. The origin of such acceleration is more interesting and more challenging issue. It is non trivial to address the cause of the late time acceleration from fundamental physics. Moreover, it is essential to incorporate a phase of inflation in early universe in the standard Big-Bang cosmology as the present observations favour such an initial phase. The Big-Bang cosmology, despite its success, has always been a concern for many who remained sceptical about its initial singularity. Emergent universe (EU) models were studied as early as in 1965 by Harrison \cite{har}. Later, Ellis \cite{ellis} studied a similar model of universe without any initial singularity.  An interesting solution was obtained by Mukherjee {\it et. al.} \cite{euorg}  where they obtained an eternally inflating solution in flat universe, which the called 'Emergent Universe', using General Relativity only and considering a non linear equation of state as below. 
\begin{equation}
\label{eos1}
p=A \rho - B \rho^{\frac{1}{2}}.
\end{equation}
It was suggested that the non linear equation of state could mimic the evolution of a universe with a mixture of three different matter energy content. The composition of the universe, they argued, would depend on the choice of the parameter $A$. Such a non linear equation of state is a special case of a more general equation $p=A \rho - B \rho^{\alpha}$. In string theory phenomenological representations of such equation of states can be found. The models based on such equation of state often interpolate between two phases of the universe \cite{fab}.  The model was later studied in different frameworks such as Brane world \cite{b1, deb}, Gauss-Bonnet gravity \cite{eugb}, Brans-Dicke theory \cite{eubd} etc. Apart from the two parameters coming from the equation of state the model involves a third parameter ($K$) as an integration constant which should be fixed by the suitable choice of initial condition. Recently, attempts were made to constrain the parameters of the original model in \cite{me1, me2, me3, pt1}. It is suggested in \cite{me3, pt1} that some of the choices in reference \cite{euorg} are permitted by present observational data. It is critical  that we have a clear idea if present observational data permits us to distinguish between different models belonging to this class. Present scheme is straightforward to explain. SNIa data tabulates distance modulus ($\mu(z)$) values obtained at different redshifts ($z$). $\mu(z)$ values are theoretically calculated for different emergent universe models. The relative difference in distance modulus values can also be obtained as $\frac{\Delta \mu}{\mu}$ for these models. The models are distinguishable only if the value of this  relative difference function ($\frac{\Delta \mu}{\mu}$) exceeds the uncertainties of SNIa observation. Also, it is interesting to investigate the constraints on the model parameters put by union compilation data which comprises of over five hundred data points. Earlier analysis were based on mostly Observed Hubble Data (OHD) with twelve data points. OHD is a collection of measured values of Hubble parameter at different redshift values from different experiments (for details see reference \cite{pt1, pt2}).  There is a particular model  where the cosmic fluid behaves as a mixture of matter, exotic matter and dark energy (for detail discussion see \cite{euorg}). If the models can be distinguished, this particular one could be an appealing candidate. 

In this particular work the above issues are addressed along with a study of late time behaviour of EU models. The plan of the paper is as follows: in section two  relevant field equations for the EU model are introduced. Data analysis based on union2 compilation of SNIa data \cite{union} is presented in section three. The possibilities of distinguishing different EU models from union2 data and study the late time behaviour of these models has been discussed. Finally, a brief discussion of findings is given in section five.

\section[]{Relevant field equations for the EU model}

Friedmann equation for a flat universe:
\begin{equation}
\label{fr1}
H^{2}=\left(\frac{\dot{a}}{a}\right)^{2}=\frac{8 \pi G \rho}{3},
\end{equation}
where $H$ is the Hubble parameter, and $a$ is the scale factor of the Universe. The conservation equation is given by
\begin{equation}
\label{csv}
\frac{d\rho}{dt}+3 H \left( p+\rho \right)=0.
\end{equation} 
Using the EOS given by eq. (\ref{eos1}) in eq.(\ref{fr1}), and eq. (\ref{csv})
\begin{eqnarray}
\label{rho1}
\rho \left(z\right)=\left(\frac{B}{A+1}\right)^{2} +\frac{2 B K}{\left(A+1\right)^2} \left(1+z\right)^{\frac{3\left(A+1\right)}{2}} \\ \nonumber +\left(\frac{K}{A+1}\right)^{2}\left(1+z\right)^{3 \left(A+1\right)}, 
\end{eqnarray}
were '$z$' is the cosmological redshift. Scale factor $a(t)$ is related to cosmological redshift ($z$): $a(t)=\frac{1}{1+z}$. The first term in the right hand side of eq.(\ref{rho1}) is a constant which can be interpreted as cosmological constant (describes dark energy). Eq. (\ref{rho1}) can be written as
\begin{equation}
\label{rho2}
\rho \left(z\right)=\rho_{1} +\rho_{2} \left(1+z\right)^{\frac{3\left(A+1\right)}{2}}+\rho_{3} \left(1+z\right)^{3 \left(A+1\right)},
\end{equation}
were $\rho_{1}=\left(\frac{B}{A+1}\right)^{2}$, $\rho_{2}=\frac{2 B K}{\left(A+1\right)^2}$, and $\rho_{3}=\left(\frac{K}{A+1}\right)^{2}$ are densities at the present epoch. The Friedmann equation (eq. \ref{fr1}) can be written in terms of redshift, and density parameter:
\begin{equation}
\label{fr2}
H^{2}\left(z\right)= H_{0}^{2}\left(\Omega_{1} +\Omega_{2} \left(1+z\right)^{\frac{3\left(A+1\right)}{2}}+\Omega_{3} \left(1+z\right)^{3 \left(A+1\right)}\right),
\end{equation}
where the density parameter is defined as $\Omega=\frac{8 \pi G \rho}{3H_{0}^{2}}=\Omega\left(A, B, K\right)$. Different composition of cosmic fluids are obtained for different values of $A$. For example, the case $A=0$ was considered in \cite{me2} and the model included dark energy, dark matter, and dust in the Universe (for details see \cite{me1}).  With $A = A_0$, eq. (\ref{fr2}) can be written as
\begin{equation}
\label{fr3}
H^{2}\left(H_{0}, B, K, z \right)= H_{0}^{2}E^{2}\left(B, K, z \right).
\end{equation}

\begin{equation}
\label{fre}
E^{2}\left(B, K, z \right)=\Omega_{\Lambda} +\Omega_{2} \left(1+z\right)^{\frac{3\left(A+1\right)}{2}}+\Omega_{3} \left(1+z\right)^{3 \left(A+1\right)},
\end{equation}
where the constant part of the DP ($\Omega_{1}$) has been replaced by a new notation $\Omega_{\Lambda}$.
\section{Analysis of the EU model with SNIa data}
In a flat universe the Hubble free luminosity distance ($D_L \equiv H_0d_L$) is defined as
\begin{equation}
\label{dl}
D_L(z)=(1+z)\int^{z}_{0}\frac{H_0}{H(z';a_1,a_2,...,a_n)}dz',
\end{equation}
where $a_1,a_2,...,a_n$ are theoretical model parameters. The distance modulus is defined as in reference\cite{nesscross} :
\begin{equation}
\label{muth}
\mu_{th}=5log_{10}(D_{L}(z)) + \mu_0,
\end{equation}
where $\mu_0=42.38-5log_{10}h$. $h$ is the dimensionless Hubble parameter at the present epoch. Consequently, a $\chi^2$ function can be defined
\begin{equation}
\label{chif}
\chi^2_{SNIa}(B, K)=\sum^{N}_{1}\frac{\left(\mu_{obs}(z_i)-\mu_{th}(z_i)\right)^2}{\sigma_i ^2},
\end{equation}
where $\mu_{obs}(z_i)$ is the observed distance modulus value at a redshift $z_i$ and $\sigma_i$ is the associated uncertainty in measurement. The union2 data set compiled in reference \cite{union} has been considered here. The above $\chi^2$ function, and the method discussed in  reference \cite{nesscross} can be used to find a $\chi^2$-fit. Findings are placed in table (\ref{sntab}).

\begin{table}
\caption[]{Best fit values of $B$ and $K$ from union2 data}
\label{sntab}
\centering
		\begin{tabular}{@{}lccc}
		\hline
		Model & $B$ & $K$ & $\chi^2_{min}$ (/d.o.f)\\
		\hline
		$A=0$ & 0.867  & 1.133 & 0.974\\
		$A=1$ & 1.491  & 0.528 & 0.985\\
		$A=1/3$ & 1.121 & 0.879 & 0.974 \\
		
		\hline	
		\end{tabular}
		
	\end{table}

\section[]{Possibilities of distinguishing different EU models with SNIa data}

\begin{figure}

\centering
\includegraphics[width=230pt,height=190pt]{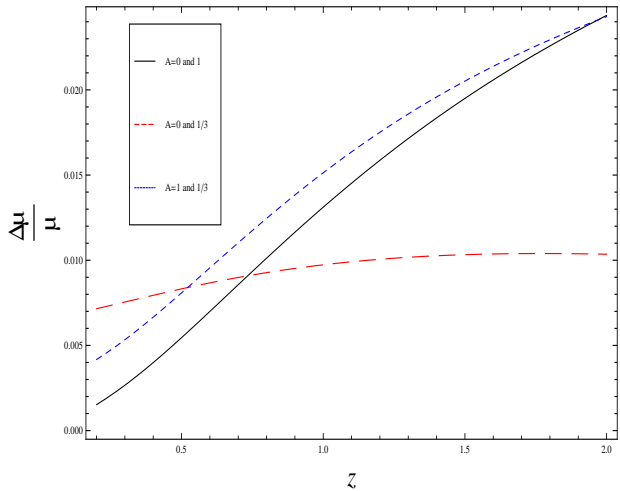}
\caption[]{(Colour Online) Evolution of difference between distance moduli for various EU models}
\label{fig::delmufig}
\end{figure}

As noted in (eq. (\ref{muth})), theoretically it is possible to obtain the distance modulus for various EU models.  Theoretical difference between the distance modulus values for two different EU models ($\Delta \mu (z)$) are calculated at different redshift points. Different EU models cannot be distinguished from SNIa observations if  $\frac{\Delta}{ \mu} (z)$ remains within the associated uncertainties in measurement of $\mu(z)$. The graphs obtained are shown in fig.(\ref{fig::delmufig}). It is seen that the $\frac{\Delta \mu}{ \mu} (z)$ values for the models with $A=0$ and $A=1$ and $A=0$ and $A=1/3$ reaches around $2\%$ or more only for $z \geq 1$. The associated uncertainties in measurement of $\mu(z)$ is around $2-4 \%$ and for redshifts $z \sim 1.5$ and higher, the uncertainty is even greater. Thus, it is not possible to distinguish between different EU models from SNIa observation . Distinguishing between EU with $A=1$ and $A=1/3$ is even more unlikely.

\subsection[]{Late time behaviour of EU models}
\label{prday}

\begin{figure}

\centering
\includegraphics[width=230pt,height=190pt]{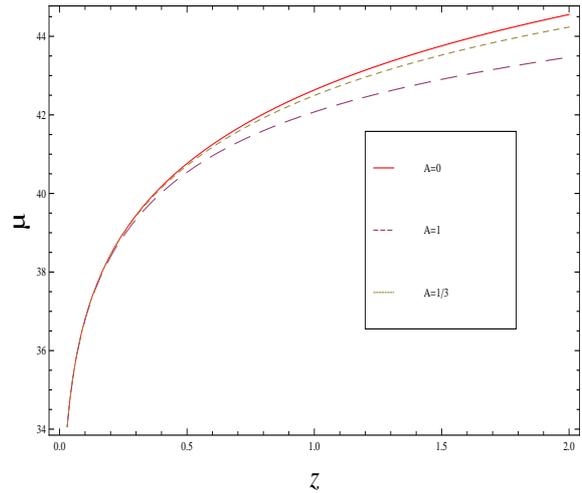}
\caption[]{(Colour Online) $\mu$ vs. $z$ curve for different EU models}
\label{fig::mufig}
\end{figure}
\begin{figure}

\centering
\includegraphics[width=230pt,height=190pt]{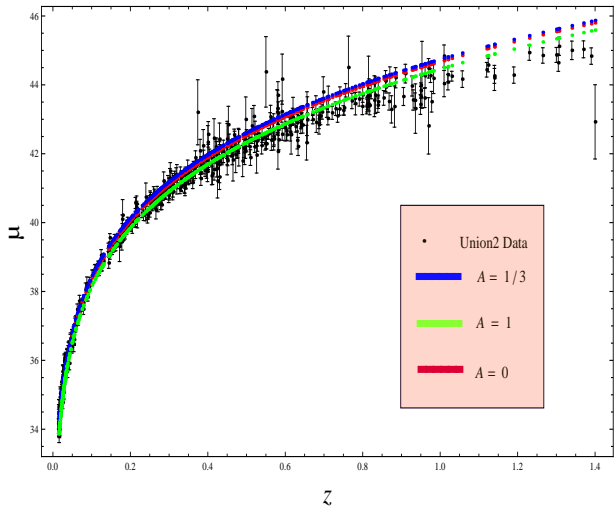}
\caption[]{(Colour Online) $\mu$ vs. $z$ curve for different EU models in late time approximation}
\label{fig::asymfig}
\end{figure}

It is seen from fig.(\ref{fig::mufig}) that the difference between different EU models fades when the present epoch is approached. This is not unexpected as these models are asymptotically de-Sitter models. Thus at late time their behaviour should be indistinguishable from one another as well as from a de Sitter universe. This can also be inferred from fig.(\ref{fig::delmufig}). The $\Delta \mu$ values for any two EU models falls within the uncertainty in the measured $\mu$ value (which is around $2-4 \%$) for lower redshifts. There is a strong possibility that at the present epoch late time behaviour of EU models is observed only. All the EU models, belonging to the class under discussion, are asymptotically de'Sitter. As given in eq. ($14$) in reference \cite{euorg} the Hubble parameter for EU is:
\begin{equation}
\label{halpha}
\centering
H=\frac{\omega \alpha e^{\alpha t}}{\beta+e^{\alpha t}},
\end{equation}
where $\beta$ is a constant $\alpha=\frac{\sqrt{3}}{2}B$, and $\omega=\frac{2}{3(A+1)}$. In late time approximation $H\approx\omega \alpha$. The $\mu$ vs. $z$ curve for different EU models along with the original union2 data is presented in (fig. (\ref{fig::asymfig})).Late time behaviour of these EU models are almost the same and typical to a de'Sitter model. These late time approximations fit union2 data reasonably well. However, as noted previously, the behaviour is typical for any de'Sitter universe. At late time the models no longer depend on the parameter $K$. Once $A$ is specified there is only one free parameter i.e., $B$.

\section{Conclusion}
A class of EU models, presented in  reference \cite{euorg}, is studied and best fit values of the the model parameters are determined from union2 compilation of SNIa data \cite{union}. More importantly, the possibilities of distinguishing different EU models from SNIa observations has also been considered. It is seen that the model with $A=1$ and $A=1/3$ can be distinguished from $A=0$ model with the data. The difference shows prominence over the uncertainty in measurement from around $z=0.5$. It has been shown that any distinction between $A=1$ and $A=1/3$ models can not be made from SNIa data as the difference remains within observational uncertainty. However, it should be noted that SNIa data becomes more uncertain at redshifts above $z=1$ and any distinction is not viable. At the present epoch the EU models cannot be distinguished from SNIa data as the difference becomes too small compared to uncertainties in the observed data. The behaviour of all EU models in the present era are typical for any asymptotically de'Sitter model. If the late time approximation of EU models are considered, it is noted that the models are independent of the parameter $K$. If $K$ is the parameter to be fixed from the initial conditions, as it is claimed in \cite{euorg}. Late time behaviour of all the EU models are independent of the initial conditions which is reasonable. The best fit values obtained here differ from the earlier works (\cite{me1, me2, me3}) and more accurate in sense that significantly more data points have been considered here. It has been shown, in a recent work on EU models (\cite{pt1}), that these models are acceptable from present observations. Present findings are in agreement with the conclusions of reference \cite{pt1}.  It would be interesting to further check these constraints growth parameter measurement which will be taken up elsewhere.

\section*{Acknowledgement}

SG is thankful to SIEM, Siliguri and IRC, University of North Bengal for providing research support.


\begin{thebibliography}{99}
\bibitem{riess} Riess et al., Astron J., 116, 1009 (1998)
\bibitem{perlm}  Perlmutter S. et al., Nature, 51, 391 (1998)
\bibitem{perlm2} Perlmutter S. et al., Astrophys. J., 517, 565 (1999)
\bibitem{perlm3} Perlmutter S. et al., Astrophys. J., 598, 102 (2003)
\bibitem{har} Harrison E. R., Mont. Not. R. Aston. Soc., 69, 137 (1967)
\bibitem{ellis} Ellis G. F. R., Maartens R., Class. Quant. Grav., 21, 223 (2004)
\bibitem{euorg} Mukherjee S. et al., Class. Quant. Grav., 23, 6927 (2006)
\bibitem{fab} Fabris J. C. et. al., Phys. Lett. A, 367, 423 (2007)
\bibitem{b1} Banerjee A., Bandyopadhyay T. and Chakraborty S., Gen.Rel.Grav., 40, 1603 (2008)
\bibitem{deb} Debnath U., Class. Quant. Grav., 25, 205019 (2008)
\bibitem{eugb} Paul B. C. and Ghose S., Gen. Rel. Grav., 42, 795 (2010)
\bibitem{eubd} del Campo S.,Herrera R., Labrana P., JCAP, 30, 0711 (2007)
\bibitem{me1} Paul B. C., Thakur, P. Ghose S., Mon. Not. R. Astron. Soc., 407, 415 (2010)
\bibitem{me2} Paul B. C., Ghose S., Thakur P., Mon. Not. R. Astron. Soc., 13, 686 (2011)
\bibitem{me3} Ghose S., Thakur P., Paul B. C., Mon. Not. R. Astron. Soc., 20, 421 (2012)
\bibitem{pt1} Thakur P., Pramana -J. Phys., 89, 27 (2017)
\bibitem{pt2} Thakur P., Pramana -J. Phys., 88, 51 (2017)
\bibitem{union} Amanullah R. et. al., Astrophys. J., 716, 712 (2010)
\bibitem{nesscross} Nesseris S., Perivolaropoulos L., JCAP, 0701, 018 (2007)





\end{thebibliography}
\end{document}